# Mechanical Properties of Minerals in Lunar and HED Meteorites from Nanoindentation Testing: Implications for Space Mining

Eloy PEÑA-ASENSIO[1,2*], Josep M. TRIGO-RODRIGUEZ[2,3], Jordi SORT[4,5], Jordi IBAÑEZ-INSA[6], and Albert RIMOLA[1]

[1] Departament de Química, Universitat Autònoma de Barcelona 08193 Bellaterra, Catalonia, Spain
[2] Institut de Ciències de l'Espai (ICE, CSIC) Campus UAB, C/ de Can Magrans s/n, 08193 Cerdanyola del Vallès, Catalonia, Spain
[3] Institut d'Estudis Espacials de Catalunya (IEEC) 08034 Barcelona, Catalonia, Spain
[4] Departament de Física, Universitat Autònoma de Barcelona, E-08193 Cerdanyola del Vallès, Spain
[5] Institució Catalana de Recerca i Estudis Avançats (ICREA), Pg. Lluís Companys 23, E-08010 Barcelona, Spain
[6] Geosciences Barcelona (GEO3BCN-CSIC), Lluís Solé i Sabarís s/n, E-08028 Barcelona, Spain
*Corresponding author. E-mail: eloy.pena@uab.cat, eloy.peas@gmail.com

**Abstract-** This study delivers an analysis of the mechanical and elemental properties of lunar meteorites DHOFAR 1084, JAH 838, NWA 11444, and HED meteorite NWA 6013. Utilizing microscale rock mechanics experiments, i.e., nanoindentation testing, this research reveals significant heterogeneity in both mechanical and elemental attributes across the mineral of the samples. Olivines, pyroxen, feldspar, and spinel demonstrate similar compositional and mechanical characteristics. Conversely, other silicate and oxide minerals display variations in the mechanical properties. Terrestrial olivines subjected to nanoindentation tests exhibit increased hardness and a higher Young's modulus compared to their lunar counterparts. A linear correlation is observed between the $H/E_r$ ratio and both plastic and elastic energies. Additionally, the alignment of mineral phases along a constant $H/E_r$ ratio suggests variations in local porosity. This study also highlights the need for further research focusing on porosity, phase insertions within the matrix, and structural orientations to refine our understanding of these mechanical characteristics. The findings have direct implications for in-situ resource utilization (ISRU) strategies and future state-of-the-art impact models. This comprehensive characterization serves as a foundational resource for future research efforts in space science and mining.

**Keywords-** Meteorites, Moon, Asteroid, ISRU, Mining, Mechanical Properties, Mineralogy

# 1. INTRODUCTION

In recent years, there has been a resurgence of interest in the exploration and colonization of the Moon and asteroids. With the increasing interest in lunar missions (**Jia et al., 2018; NASA, 2020; Mitrofanov, 2021; Bhardwaj, 2021**), it is important to understand the properties of the materials available on the lunar surface for in-situ resource utilization (ISRU) and manufacturing (**Anand et al., 2012; Crawford, 2015; Naser, 2019; Duffard et al., 2021**). The extension of space exploration endeavors to include near-Earth objects necessitates an enhanced understanding of both asteroid mining techniques and science (**Ross, 2001; Anthony, 2018; Zacny, 2013; Andrews, 2015**).

Lunar meteorites, also known as lunaites, are fragments of the Moon ejected by impacts that have subsequently fallen to Earth and provide a unique opportunity to study the composition and properties of lunar materials. Unlike the samples returned from the Apollo missions, which were collected by astronauts from limited locations, lunar meteorites are fragments of the Moon of unknown but, more importantly, random origin. These meteorites are excellent samples to study a wider range of lunar materials that were not screened during the astronaut extravehicular activities. Lunar meteorites have been found in various locations around the world and have been extensively studied using a variety of techniques, including petrographic analysis, X-ray diffraction, and mass spectrometry, among others. These studies have revealed a wealth of information about the mineralogy, chemistry, and geologic history of the Moon (**Korotev, 2005**). Similarly, despite the significant contributions of the Stardust (**Brownlee et al., 2006**), Hayabusa (**Yano et al., 2006**), Hayabusa2 (**Watanabe et al., 2019**), and OSIRIS-Rex (**Lauretta et al., 2019**) sample return missions, meteorites are the primary source of samples essential for inferring asteroidal and cometary properties.

Indentation and tensile testing are common methods for measuring the mechanical properties of minerals, including hardness and Young's modulus. However, due to the rarity of lunar meteorites, large-scale destructive experiments cannot be performed as they must be preserved for their scientific value. As a result, researchers have turned to microscale rock mechanics experiments as a nearly non-destructive method for characterizing the mechanical properties of meteorites (**Moyano-Cambero et al., 2017; Tanbakouei et al., 2019; Wheeler, 2021; Zhang et al., 2022; Huang, 2023; Nie et al., 2023; Rabbi, 2023**). The nanoindentation technique allows probing the physical properties of materials at the nanoscale and microscale. This is particularly important for the investigation of meteorites because: i) it can provide valuable insight into the properties of the constituent minerals, which do not need to be exactly equivalent to their terrestrial counterparts due to, for instance, compositional differences and different shock degrees; ii) knowledge of the mechanical properties of the minerals, combined with appropriate modeling of the behavior under different regimes (elastic, plastic, etc.), may allow to infer the mechanical properties of asteroidal and lunar surface rocks in different scenarios relevant for ISRU; iii) from a more fundamental point of view, nanoindentation may be useful to gain information about how the shock history of the meteorites may affect their macro- and micro-structural properties and mechanical behavior; and iv) in composite rocks, how the individual mechanical properties of each constituent mineral correlate with the overall mechanical performance is of interest, for example, for mining prospections.

In mining engineering and geomechanical analysis, differentiating between the mechanical properties of minerals and rocks is fundamental. Minerals' mechanical properties, while insightful, do not directly translate to rocks' macroscopic behaviors. This necessitates methods to extrapolate microscale (mineral-level) properties to macroscale (rock-level) phenomena. **Xu *et al.* (2023)** exemplify this by studying thermally induced microcracks in granite, analyzing mineral heterogeneity and thermal stress effects on granite's macroscale mechanical integrity using high-temperature microscopy and Accurate Grain-Based Modeling (AGBM). Conversely, **Tang *et al.* (2023**) employ microscale rock mechanics experiments and AGBM to determine asteroid rocks' Young's modulus, bridging mineral and interphase-level analyses to a comprehensive macroscale understanding.

In relation to ISRU, it is important to emphasize that lunar materials have unique mechanical properties because of their composition and structure. Terrestrial rocks are composed primarily of silica- and silicate-based minerals such as quartz, feldspar, and mica, along with notable elemental constituents such as iron, aluminum, calcium, and magnesium. Lunar rocks have a similar silicate composition but exhibit distinct mineralogical variations (**Papike, Taylor, & Simon, 1991**). In addition, lunar rocks have a higher abundance of elements such as titanium, thorium, and rare earth elements. Besides compositional considerations, Earth's materials have undergone a variety of recent geologic processes, including erosion, weathering, and tectonic activity. On the contrary, lunar material, exposed to the vacuum of space, has experienced limited influence from these geology processes. Nevertheless, lunar rocks have been subjected to the so-called *space weathering*: impact events caused by asteroids, meteoroids, and micrometeoroids, resulting in fractures and subsequent physical changes due to extreme temperature and pressure variations (**Hapke, 2001**).

Significant progress has been made in the development of lunar simulants, which are synthetic materials designed to mimic the properties of lunar materials (**McKay et al., 1994; Kanamori et al., 1998; Zheng et al., 2009; Alberquilla et al., 2022**). These simulants play a critical role in various scientific and engineering research activities related to lunar exploration and colonization. However, it remains essential to study lunar samples rather than relying solely on analogs, as they cannot perfectly replicate the precise characteristics and complexities of real materials. Lunar simulants often approximate certain properties such as composition, grain size, and density, but they may not accurately capture the intricate details and variations found in real samples from the Moon. Therefore, it is necessary to study the mechanical properties of real samples to design appropriate manufacturing techniques and equipment for the lunar environment. By characterizing the mechanical behavior of lunar materials, we can develop suitable processes and tools for lunar exploration and colonization, and ultimately improve our understanding of the Moon's resources and potential for human exploration and habitation. Similarly, analyzing materials from asteroids follows the same principle, contributing to our comprehension of their utility in space exploration and science.

In addition, the mechanical characterization of extraterrestrial material components is essential for future state-of-the-art impact modeling. Impact events are a ubiquitous process on the lunar and asteroid surfaces, and understanding the mechanics of these events is important for several fields, like geomorphological characterization of craters. While macroscopic mechanical properties are essential for understanding the behavior of materials under impacts and other extreme conditions, incorporating the micromechanical properties of the different phases of a material can provide a more accurate representation of the overall material's behavior, particularly when it is composed of several phases or minerals (**Melosh, 1989; Davison, Collins, & Bland, 2016**). Simulations that incorporate the micro- and nanomechanical properties of different phases can provide a more detailed representation of how macroscopic composite materials respond to impact and other extreme conditions. This can be particularly important for materials that are complex and heterogeneous, such as lunar meteorites, where the behavior of different phases can vary significantly due to differences in their composition, grain size, and orientation.

In this study, we selected three lunar meteorites and one howardite–eucrite–diogenite (HED) meteorite for nanoindentation tests, aiming to generate valuable data pertinent to both lunar and asteroidal mining and science. In the next section, we will introduce the samples studied. Then, we will describe the results obtained for the main rock-forming minerals that have been tested. Finally, we will compare our results with previous data published in the literature. Our main goal is to exemplify the intrinsic value of a systematic study of meteorites using nanoindentation to infer the main mechanical properties of extraterrestrial minerals for ISRU.

## 2. SAMPLES AND EXPERIMENTAL METHODS

The thin section of the meteorites analyzed in this work (DHOFAR 1084, JAH 838, NWA 11444, and NWA 6013) belong to the Meteorite Collection of the Institute of Space Sciences (CSIC). The four samples are classified in the Meteoritical Bulletin Database, and their main characteristics are compiled in **Table 1**.

Table 1. Classification of the 4 samples analyzed in this work. Data extracted from the Meteoritical Bulletin Database. Note that NWA 6013 is not lunar.

| Name | Classification | Fayalite (mol %) | Ferrosilite (mol %) |
|---|---|---|---|
| DHOFAR 1084 | Feldespathic | 43 | 30 |
| JAH 838 | Mingled regolith breccia | 39.7 | 55.6 |
| NWA 11444 | Melt breccia | 37.1 | 35.1 |
| NWA 6013 | Hartzburgitic diogenite | 29.5 | 24 |

DHOFAR 1084 is a lunar meteorite found in 2001 in the Dhofar region of Oman, near Zufar. This meteorite has been classified as an average feldspathic lunar meteorite. Near this meteorite, DHOFAR 490 was also discovered, and both were isolated from other lunar rocks. However, it has still to be determined if they constitute a pairing. **Fig. 1** shows the thin section obtained from DHOFAR 1084.

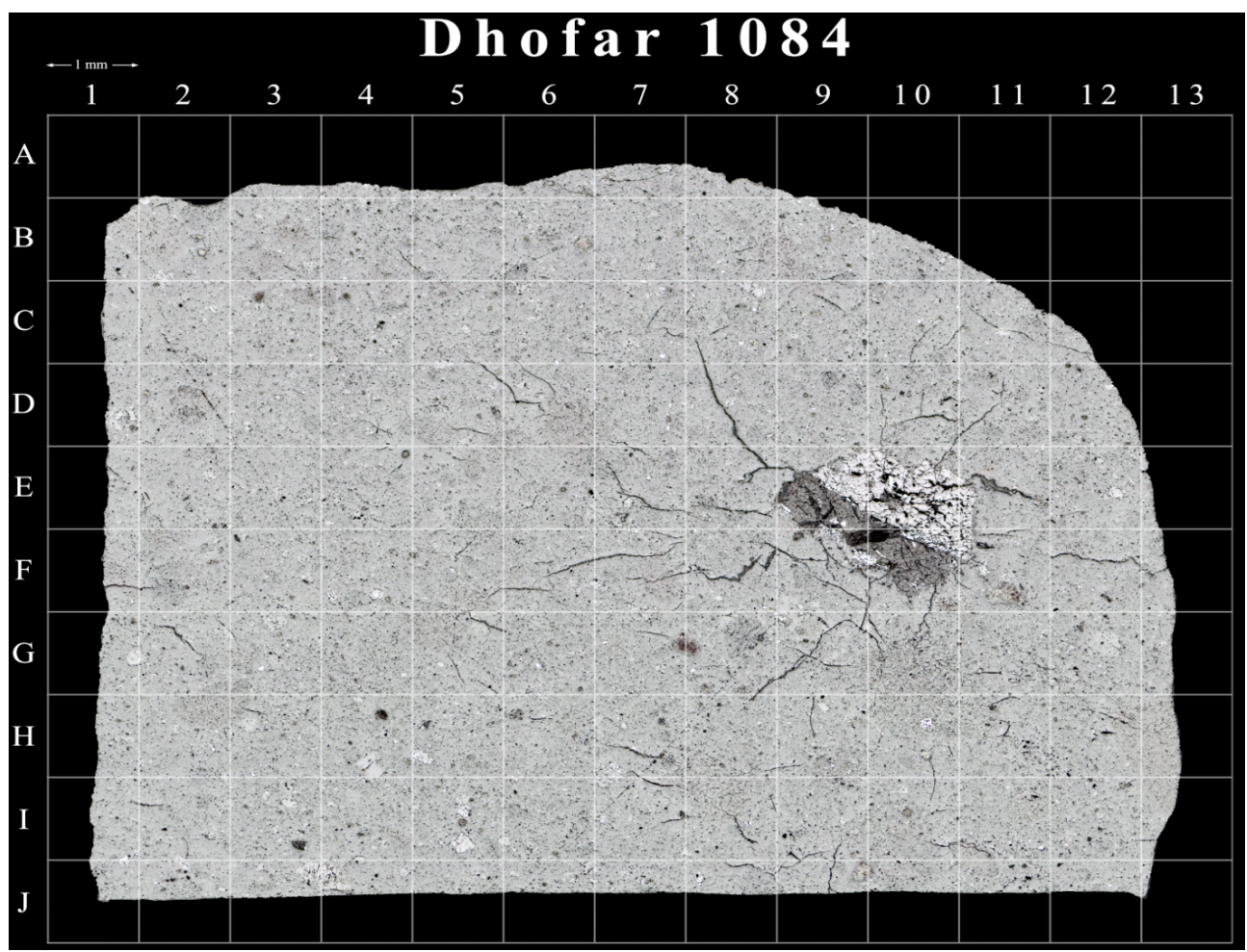

**Figure 1.** Mosaic of the thin section of lunar meteorite DHOFAR 1084.

Jiddat al Harasis 838 (JAH 838) is a lunar meteorite found in Oman during a desert expedition in 2003, 28 km south of the Al Ghaftain. It is classified as a mingled regolith breccia and has a presence of mare and KREEPy material (potassium, rare-earth elements, and phosphorus), HASP (alumina–silica poor), and chondritic material., **Fig. 2** shows the thin section of JAH 838.

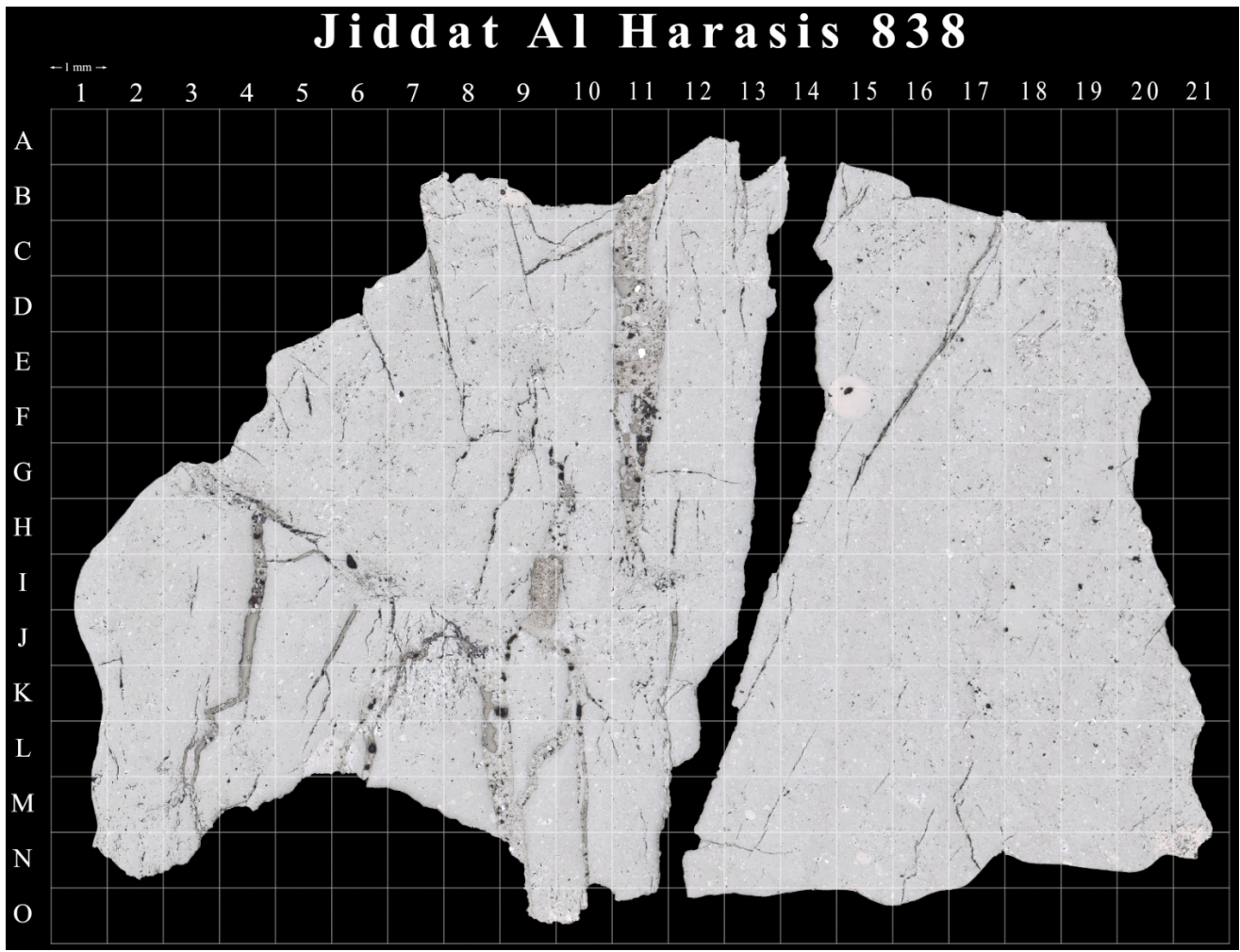

**Figure 2.** Mosaic of the thin section of lunar meteorite JAH 838.

Northwest Africa 11444 (NWA 11444) is a lunar meteorite discovered in 2017 and collected in an unknown location in Mauritania. It is classified as a polymictic anorthositic breccia, which means that it is composed of fragments of anorthosite, a type of rock that is rich in the mineral plagioclase, fused together by the heat generated by a large impact event on the Moon's surface. **Fig. 3** shows the thin section obtained from NWA 11444.

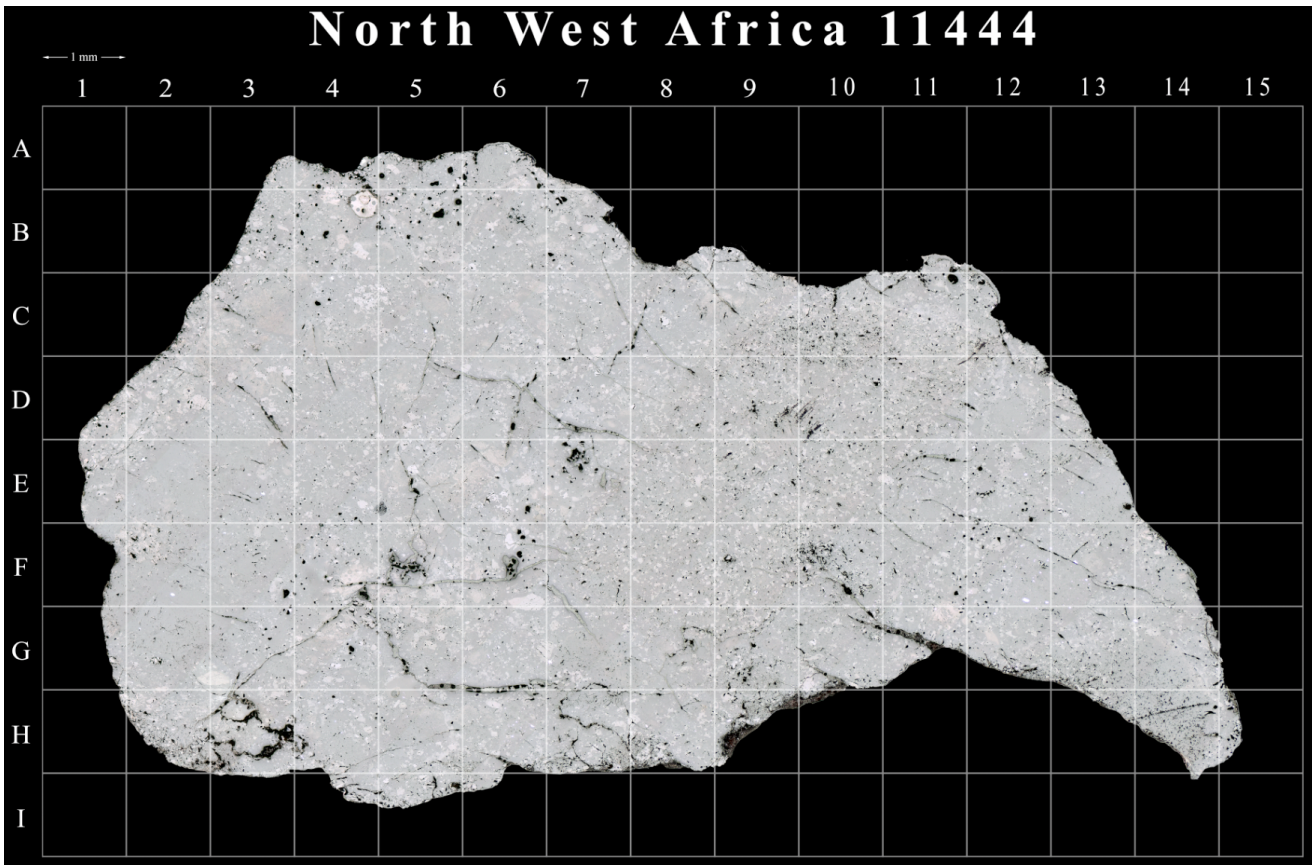

**Figure 3.** Mosaic of the thin section of lunar meteorite NWA 11444.

Northwest Africa 6013 (NWA 6013) is a meteorite discovered in 2008 in Northwest Africa. It is classified as a Hartzburgitic diogenite, that is, HED achondrite, a type of ultramafic rock with a coarse-grained texture that contains subunits dominated by olivine and pyroxene with approximately equal abundances of 50 vol% each. **Fig. 4** shows the thin section of NWA 6013.

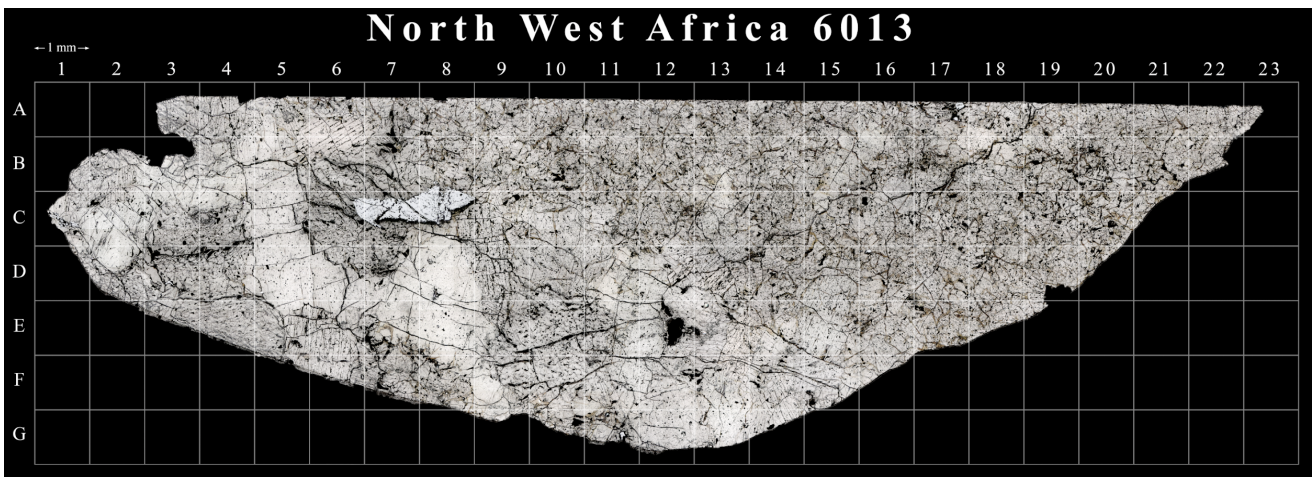

**Figure 4.** Mosaic of the thin section of meteorite NWA 6013.

A petrographic optical microscope, Zeiss Scope Axio, was used to examine the thin sections of the samples in both reflected and transmitted light modes at ×50, ×100, and ×250 magnification. To identify and name the various features and components, high-resolution mosaics were created, and a grid was superimposed on the mosaics. The creation of a grid mosaic was essential to facilitate the process of the identification of the Regions of Interest (ROIs) to perform the nanoindentation and mineral identification, as it allowed precise placement of the indenter tip on specific minerals and provided a visual reference for their location within the sample. This approach ensured accurate characterization of the mechanical properties of individual minerals within the lunar meteorites.

We used scanning electron microscopy (SEM) with energy dispersive X-ray analysis (EDX) from the Catalan Institute of Nanoscience and Nanotechnology (ICN2), specifically the FEI Quanta 650 FEG in low vacuum backscattered electron mode (BSED), to examine the thin sections of the samples. An Inca 250 SSD Xmax20 EDS detector was used for elemental analysis. The EDS detector is equipped with Peltier cooling and has an active area of 20 mm$^2$. Various magnifications were used to examine selected areas of the samples, from which we obtained the corresponding EDX spectra. This approach allowed a detailed analysis of the mineralogical composition of the selected ROIs and an overview of the elemental distribution within the sections. The information obtained from these analyses allowed us to identify the indented minerals.

The mechanical properties of the minerals were obtained using the NHT2 Anton Paar nanoindentation instrument available at the Autonomous University of Barcelona (UAB), which uses a Berkovich pyramidal diamond tip. In the process of nanoindentation, controlled force is precisely applied to specific and localized areas of a sample using a diamond indenter. The indenter exerts pressure on the surface as the force is gradually increased up to a predetermined maximum value. Subsequently, after the loading phase, the force is systematically reduced back to zero, and the sample's surface retracts to some extent due to its elasticity. Throughout this entire load-unload cycle, the penetration depth is measured. By analyzing the load-depth curves derived from this process, valuable insights into deformation mechanisms (both elastic and plastic) and elastic recovery are obtained.

To evaluate average mechanical properties, an array of 6-12 indentations was performed for each mineral with a maximum applied force of 25 mN. The thermal drift during nanoindentation was carefully kept below 0.05 nm/s. Precise corrections for contact area (calibrated with a fused silica sample), initial indentation depth, and instrument compliance were made (**Fischer-Cripps & Nicholson 2004**). Hardness ($H$) and reduced Young's modulus ($E_r$) values were obtained from the load-displacement curves following the method of **Oliver & Pharr (1992)** as depicted in **Fig. 5**.

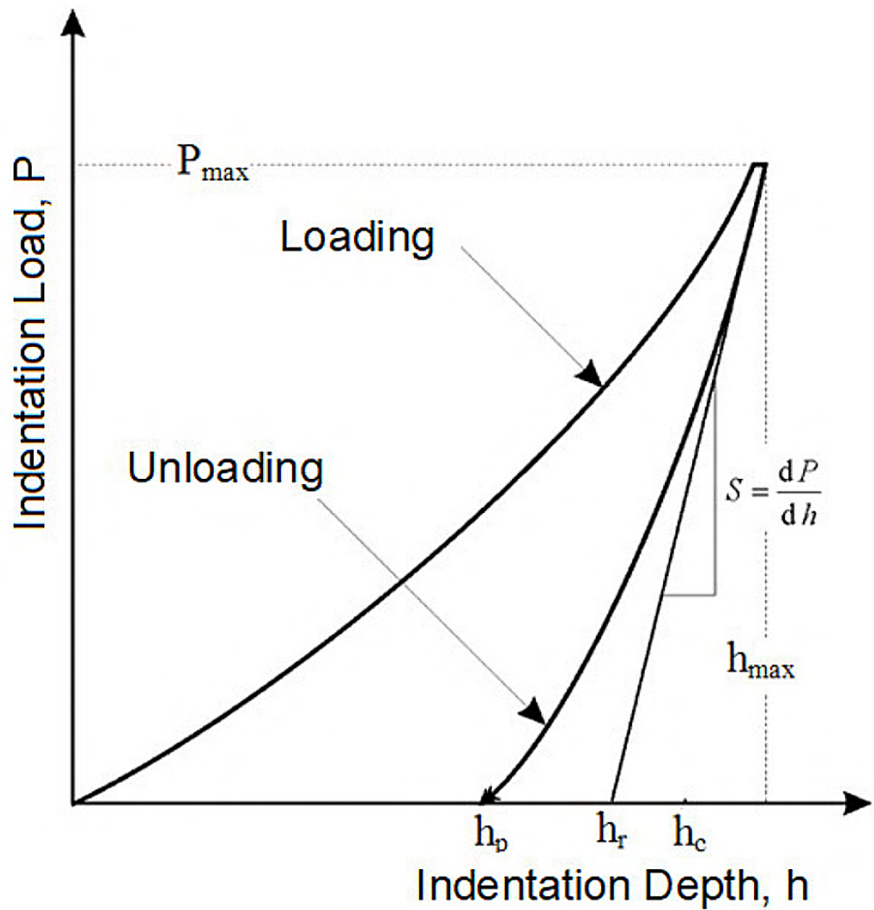

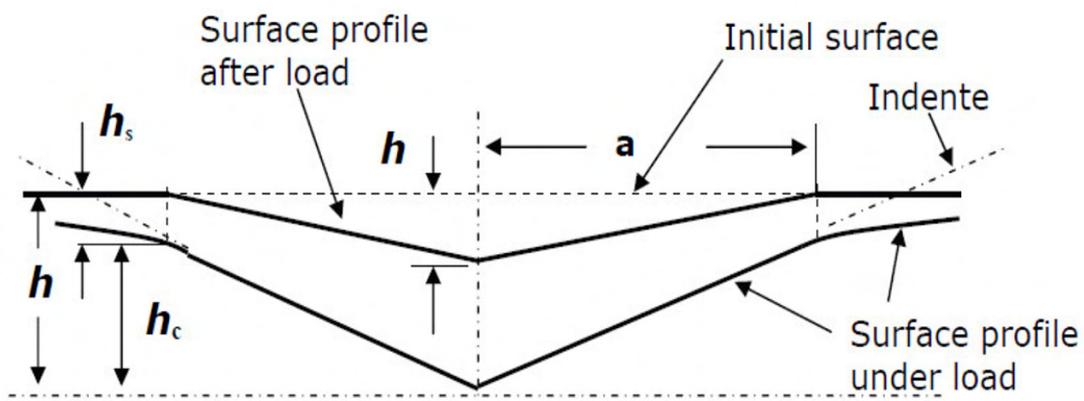

**Figure 5**. Top: Load-displacement curve, showing the parameters used in the Oliver and Pharr method. Bottom: cross-section of an indentation. Adapted from **Ovsik et al. (2021)**.

The Young's modulus refers to a fundamental mechanical property that quantifies the stiffness or elasticity of a material., It measures the ability of a material to resist deformation under an applied force. On the other hand, the reduced Young's modulus (which is, in fact, the elastic property measured during nanoindentation) refers to a modified or adjusted version of Young's modulus, that considers convoluted contributions from the sample and the indenter tip (i.e., two-body contact):

$$\frac{1}{E_r} = \frac{1-v^2}{E} - \frac{1-v_i^2}{E_i} \qquad (1)$$

The reduced modulus considers the elastic displacements occurring in both the specimen, characterized by Young's modulus $E$ and Poisson's ratio $v$, and the diamond indenter, which has Young's modulus and Poisson ratio equal to $E_i$ = 1,140 GPa and $v_i$ = 0.07, respectively. The hardness $H$ is computed as:

$$H = \frac{P_{max}}{A} \qquad (1)$$

where $P_{max}$ is the maximum load applied and $A$ is the projected contact area on the horizontal plane. The parameter $H/E_r$ merits consideration as it provides an indirect measure of wear resistance (**Pellicer et al., 2021**).

Elastic recovery values were obtained from the ratio between elastic and total indentation energies ($W_e/W_t$). $W_e$ was determined from the area between the unloading curve and the displacement axis, whereas $W_t$ was estimated from the area between the loading curve and the displacement axis. The plastic behavior was characterized analogously. The parameter $W_p/W_t$ is often referred to as the plasticity index.

## 3. RESULTS

To gain valuable insights into the nature of lunar meteorites, we conducted a comprehensive analysis of their mechanical properties, identifying the main mineral of each tested region. **Table 2** summarizes the mechanical properties obtained from nanoindentation testing for each of the studied lunar meteorites. We examined the atomic percentage composition of these lunar meteorites using SEM EDX measurements. **Table 3** displays the atomic composition results, offering insights into the elemental composition of each indented mineral. Note that silicates that cannot be distinctly classified as pyroxenes, olivines, or feldspars, and which present ambiguous characteristics under SEM analysis, are categorized as 'other silicate'.

**Figure 6** presents four distinct examples of indentation tests conducted on various minerals found within each of the lunar meteorite samples, accompanied by their respective optical microscope images. The differences in the appearance of each indentation result from variations in illumination and reflection conditions. In the top row, corresponding to diopside/augite in the meteorite DHOFAR 1084, the load curves usually exhibit small and abrupt flat regions, potentially suggestive of porosity within the material or weak inclusions of phase within the matrix. Conversely, in the third row (pigeonite/enstatite in NWA 11444), several measurements had to be excluded due to the occurrence of sudden drops in the indentation curves, suggestive of cracking or pore collapse. Nonetheless, the figure also shows the case of an outlier curve, which coincides with a fracture occurrence within the material.

**Table 2.** Compilation of the mechanical properties obtained for each mineral and meteorite from nanoindentation testing. The last column indicates the number of indentations performed at each test.

| Meteorite | Cell | Mineral | Group | H (GPa) | $E_r$ (GPa) | $H/E_r$ | $W_e/W_t$ | $W_p/W_t$ | # |
|---|---|---|---|---|---|---|---|---|---|
| DHO 1084 | I11 | Anorthite | Feldspar | 9.6±0.9 | 92±7 | 0.105±0.013 | 0.57±0.04 | 0.43±0.05 | 12 |
| DHO 1084 | F11 | Anorthite (Oliv.) | Feldspar | 9.2±0.5 | 93±5 | 0.099±0.008 | 0.557±0.033 | 0.443±0.027 | 9 |
| DHO 1084 | E11 | Anorthite | Feldspar | 8.8±0.2 | 99±7 | 0.090±0.007 | 0.50±0.05 | 0.50±0.06 | 6 |
| DHO 1084 | I12 | Diopside/Augite | Pyroxene | 10.7±0.7 | 146±3 | 0.073±0.005 | 0.431±0.019 | 0.569±0.031 | 8 |
| DHO 1084 | H10 | | | 14.3±3.4 | 149±18 | 0.096±0.026 | 0.50±0.08 | 0.50±0.15 | 7 |
| DHO 1084 | F10 | Other silicate | Other | 6.7±0.9 | 66±5 | 0.102±0.016 | 0.60±0.06 | 0.40±0.07 | 8 |
| DHO 1084 | A8 | Titanomagnetite | Spinel | 10.6±0.4 | 154±2 | 0.0685±0.0025 | 0.392±0.021 | 0.608±0.035 | 6 |
| JAH 838 | L6 | Anorthite | Feldspar | 9.3±0.5 | 90±5 | 0.103±0.008 | 0.583±0.032 | 0.417±0.023 | 8 |
| JAH 838 | L6 | Forsterite | Olivine | 10.2±1.9 | 139±15 | 0.073±0.016 | 0.40±0.05 | 0.60±0.11 | 8 |
| JAH 838 | L6 | Calcite | Carbonate | 2.8±0.2 | 74±8 | 0.038±0.005 | 0.27±0.04 | 0.73±0.07 | 8 |
| JAH 838 | L6 | Forsterite | Olivine | 8.8±2.4 | 113±21 | 0.078±0.026 | 0.42±0.09 | 0.58±0.15 | 9 |
| JAH 838 | M3 | Anorthite | Feldspar | 7.9±0.5 | 79±2 | 0.099±0.007 | 0.56±0.02 | 0.44±0.04 | 5 |
| JAH 838 | M3 | Anorthite | Feldspar | 7.4±1 | 83±5 | 0.088±0.013 | 0.54±0.08 | 0.46±0.09 | 8 |
| JAH 838 | N12 | Anorthite (Carb.) | Feldspar | 9.5±0.6 | 86±3 | 0.110±0.008 | 0.591±0.022 | 0.409±0.027 | 8 |
| JAH 838 | N12 | Pigeonite | Pyroxene | 11±1 | 134±6 | 0.082±0.008 | 0.45±0.02 | 0.55±0.005 | 6 |
| JAH 838 | N12 | Titanomagnetite | Spinel | 9.6±0.9 | 116±3 | 0.083±0.008 | 0.45±0.03 | 0.55±0.07 | 6 |
| JAH 838 | B9 | Anorthite | Feldspar | 10.6±0.3 | 88±1 | 0.121±0.004 | 0.589±0.0016 | 0.411±0.019 | 8 |
| JAH 838 | G11 | Other silicate | Other | 13.8±0.3 | 90±1 | 0.1525±0.0034 | 0.747±0.008 | 0.253±0.006 | 9 |
| NWA 11444 | D2 | Pige./Enst. | Pyroxene | 10.2±0.6 | 155±5 | 0.066±0.005 | 0.403±0.021 | 0.597±0.028 | 6 |
| NWA 11444 | C3 | Forsterite | Olivine | 11.7±1.1 | 156±13 | 0.075±0.010 | 0.41±0.04 | 0.59±0.07 | 7 |
| NWA 11444 | B6 | Pigeonite | Pyroxene | 12.1±0.3 | 148±1 | 0.0817±0.0023 | 0.447±0.007 | 0.553±0.008 | 8 |
| NWA 11444 | C8 | Diopside | Pyroxene | 12.2±0.7 | 162±5 | 0.075±0.005 | 0.4±0.3 | 0.6±0.4 | 6 |
| NWA 11444 | C8 | Pige./Enst. | Pyroxene | 12.3±0.5 | 147±5 | 0.084±0.004 | 0.447±0.024 | 0.553±0.023 | 7 |
| NWA 11444 | C3 | Calcite | Carbonate | 2.4±0.2 | 63±2 | 0.0377±0.0031 | 0.27±0.02 | 0.73±0.06 | 9 |
| NWA 11444 | C2 | Ilmenite | Oxide | 9.5±1 | 134±12 | 0.071±0.010 | 0.41±0.04 | 0.59±0.06 | 9 |
| NWA 11444 | D2 | Pige./Enst. | Pyroxene | 9.9±0.9 | 139±8 | 0.071±0.008 | 0.41±0.04 | 0.59±0.09 | 7 |
| NWA 11444 | D2 | Pige./Enst. | Pyroxene | 10±0.6 | 135±6 | 0.074±0.006 | 0.44±0.03 | 0.56±0.04 | 8 |
| NWA 11444 | D2 | Anorthite | Feldspar | 9±0.6 | 92±9 | 0.097±0.012 | 0.56±0.05 | 0.44±0.04 | 7 |
| NWA 11444 | C8 | Forsterite | Olivine | 12.5±0.5 | 174±4 | 0.072±0.0031 | 0.402±0.012 | 0.598±0.019 | 9 |
| NWA 11444 | E4 | Other silicate | Other | 4.9±1.3 | 38±5 | 0.13±0.04 | 0.67±0.08 | 0.33±0.01 | 7 |
| NWA 6013 | C16 | Chromite | Oxide | 17.3±0.3 | 147±2 | 0.1178±0.0023 | 0.583±0.010 | 0.417±0.014 | 9 |
| NWA 6013 | C15 | Troilite | Sulfide | 12.7±0.3 | 147±2 | 0.0864±0.0025 | 0.48±0.01 | 0.520±0.021 | 7 |
| NWA 6013 | F14 | Labradorite | Feldspar | 10.6±0.3 | 88±1 | 0.120±0.004 | 0.601±0.013 | 0.399±0.016 | 9 |
| NWA 6013 | F14 | Pigeonite | Pyroxene | 12.6±0.3 | 134±2 | 0.0946±0.0026 | 0.497±0.008 | 0.503±0.015 | 9 |

**Table 3.** Compilation of the atomic percentage composition from each nanoindentation performed. The last column indicates the number of measurements performed at each test.

| Meteorite | Cell | Mineral | Group | O (%) | Na (%) | Mg (%) | Al (%) | Si (%) | S (%) | Ca (%) | Ti (%) | Cr (%) | Fe (%) | Other (%) | # |
|---|---|---|---|---|---|---|---|---|---|---|---|---|---|---|---|
| DHO 1084 | I11 | Anorthite | Feldspar | 61.45±0.48 | 0.28±0.05 | 2.22±0.55 | 9.15±0.57 | 13.17±0.47 | 0.02±0.04 | 4.59±0.14 | 0.12±0.21 | | 1.34±0.39 | 7.66±2.9 | 5 |
| DHO 1084 | I13 | Anorthite (Oliv.) | Feldspar | 63.54±0.48 | 0.52±0.25 | 2.06±0.52 | 9.56±0.36 | 13.52±0.38 | 0.18±0.02 | 4.77±0.27 | | | 1.31±0.28 | 4.56±2.55 | 4 |
| DHO 1084 | I14 | Anorthite | Feldspar | 64.15±0.6 | 0.24±0.03 | 2.61±1.11 | 8.84±1.19 | 13.92±0.16 | 0.18±0.02 | 4.76±0.66 | | | 1.92±0.95 | 3.39±4.72 | 3 |
| DHO 1084 | I15 | Diopside/Augite | Pyroxene | 61.3±0.52 | | 5.98±0.16 | 3.31±0.04 | 15.66±0.26 | | 5.59±0.35 | 0.21±0.02 | 0.16±0.01 | 1.98±0.18 | 5.79±1.53 | 5 |
| DHO 1084 | I16 | | | 60.51±0.43 | | 2.56±0.15 | 8.82±0.24 | 4.35±0.35 | | 1.24±0.04 | 0.41±0.04 | 8.78±0.38 | 6.64±0.4 | 6.68±2.04 | 5 |
| DHO 1084 | I17 | Other silicate | Other | 61.6±0.45 | 0.48±0.05 | 1.35±0.28 | 3.16±0.36 | 19.52±0.81 | | 0.68±0.02 | 0.12±0.02 | | 1.45±0.15 | 11.65±2.14 | 5 |
| DHO 1084 | I19 | Titanomagnetite | Spinel | 63.36±0.72 | | 1.45±0.09 | 0.98±0.02 | 4.53±0.29 | | 0.75±0.08 | 11.75±0.21 | 0.09±0.03 | 11.94±0.22 | 5.17±1.66 | 4 |
| JAH 838 | I20 | Anorthite | Feldspar | 58.5±0.44 | 0.18±0.04 | 0.56±0.15 | 9.96±0.2 | 11.41±0.02 | | 5.29±0.07 | 0.5±0.1 | | | 13.6±1.03 | 3 |
| JAH 838 | I21 | Forsterite | Olivine | 58.73±0.32 | | 8.77±0.15 | 2.17±0.05 | 9.95±0.12 | | 1.86±0.07 | | | 6.06±0.11 | 12.47±0.83 | 4 |
| JAH 838 | I22 | Calcite | Carbonate | 57.6±0.16 | | 1.05±0.06 | 2.49±0.06 | 3.45±0.14 | | 17.07±0.16 | | | 0.69±0.06 | 17.66±0.64 | 5 |
| JAH 838 | I23 | Forsterite | Olivine | 59.36±0.21 | | 9.52±0.06 | 2.26±0.11 | 10.69±0.2 | | 1.54±0.03 | | | 6.53±0.14 | 10.11±0.74 | 6 |
| JAH 838 | I24 | Anorthite | Feldspar | 55.32±0.25 | 0.15±0.02 | 0.41±0.09 | 9.05±0.14 | 10.1±0.06 | | 4.86±0.17 | | | 0.16±0 | 19.95±0.73 | 3 |
| JAH 838 | I25 | Anorthite | Feldspar | 58.19±0.4 | 0.15±0.07 | 1.07±0.66 | 9.01±0.84 | 11.16±0.71 | | 5.37±0.58 | 0.02±0.03 | | 0.44±0.2 | 14.6±3.5 | 6 |
| JAH 838 | I26 | Anorthite (Carb.) | Feldspar | 49.39±0.57 | 0.14±0.02 | 0.29±0.02 | 7.82±0.15 | 8.47±0.26 | | 3.81±0.09 | | | 0.25±0.05 | 29.84±1.16 | 4 |
| JAH 838 | I27 | Pigeonite | Pyroxene | 52.38±0.26 | | 6.02±0.11 | 2.31±0.14 | 7.94±0.23 | | 0.99±0.04 | | | 5.14±0.11 | 25.23±0.9 | 4 |
| JAH 838 | I28 | Titanomagnetite | Spinel | 51.17±0.64 | | 1.2±0.06 | 3.79±0.08 | 3.65±0.07 | | 1.31±0.03 | 5.01±0.12 | 1.66±0.04 | 9.75±0.33 | 22.46±1.37 | 5 |
| JAH 838 | I29 | Anorthite | Feldspar | 45±5.21 | | 2.24±0.1 | 5.06±0.25 | 8.72±0.31 | | 2.8±0.19 | | | 0.91±0.13 | 35.28±6.19 | 6 |
| JAH 838 | I30 | Other silicate | Other | 60.21±0.11 | | 0.21±0.01 | 0.99±0.04 | 20.18±0.43 | | 1.19±0.04 | | | 0.11±0.01 | 17.11±0.64 | 3 |
| NWA 11444 | I31 | Pige./Enst. | Pyroxene | 61.72±1.37 | | 5.66±0.14 | 2.1±0.07 | 15.52±0.76 | | 5.58±0.44 | 0.1±0.02 | 0.18±0.02 | 2.74±0.22 | 6.42±3.04 | 4 |
| NWA 11444 | I32 | Forsterite | Olivine | 58.81±0.47 | | 17.17±0.26 | 2.87±0.08 | 12.15±0.21 | 0.04±0.07 | 1.24±0.03 | | | 1.71±0.04 | 6±1.16 | 5 |
| NWA 11444 | I33 | Pigeonite | Pyroxene | 57.54±1.03 | | 5.66±0.19 | 2.55±0.05 | 13.86±0.34 | | 1.96±0.14 | 0.08±0.01 | | 4.73±0.18 | 13.64±1.94 | 4 |
| NWA 11444 | I34 | Diopside | Pyroxene | 58.64±0.95 | | 6.27±0.29 | 3.66±0.13 | 15.36±0.18 | | 6.22±0.44 | 0.32±0.01 | 0.15±0.01 | 1.28±0.12 | 8.11±2.13 | 4 |
| NWA 11444 | I35 | Pige./Enst. | Pyroxene | 58.28±0.24 | | 13.97±0.27 | 2.92±0.13 | 11.93±0.18 | | 1.26±0.07 | | | 3.7±0.11 | 7.95±0.99 | 4 |
| NWA 11444 | I36 | Calcite | Carbonate | 57.73±1.63 | | 0.77±0.17 | 2.46±0.09 | 2.89±0.16 | | 13.81±0.72 | | | 0.22±0.02 | 22.13±2.79 | 4 |
| NWA 11444 | I37 | Ilmenite | Oxide | 56.19±0.53 | | 1.68±0.08 | 3.38±0.16 | 3.75±0.13 | | 1.32±0.07 | 9.89±0.19 | | 8.61±0.15 | 15.18±1.29 | 4 |
| NWA 11444 | I38 | Pige./Enst. | Pyroxene | 61.67±0.23 | | 6.46±0.15 | 3.26±0.12 | 15.73±0.21 | 0.12±0.09 | 2.42±0.22 | 0.08±0.04 | | 5.16±0.11 | 5.09±1.18 | 5 |
| NWA 11444 | I39 | Pige./Enst. | Pyroxene | 59.46±1.08 | | 5.27±0.29 | 3.37±0.43 | 15.41±0.31 | | 4.86±1.23 | 0.12±0.03 | 0.07±0.02 | 3.81±0.95 | 7.65±4.33 | 6 |
| NWA 11444 | I40 | Anorthite | Feldspar | 61.16±1.44 | 0.17±0.11 | 1.31±0.83 | 9.86±1.69 | 12.26±1.08 | 0.04±0.06 | 5.71±0.54 | | | 0.73±0.32 | 8.77±6.08 | 4 |
| NWA 11444 | I41 | Forsterite | Olivine | 58.9±0.48 | | 10.81±0.11 | 2.48±0.04 | 12.15±0.02 | | 1.02±0 | | | 7.97±0.06 | 6.68±0.72 | 3 |
| NWA 11444 | I42 | Other silicate | Other | 66.62±0.19 | | 0.5±0.05 | 1.5±0.03 | 26.48±0.06 | | 0.85±0.02 | | | 0.16±0.02 | 3.9±0.37 | 4 |
| NWA 6013 | I43 | Chromite | Oxide | 60.04 | 0.00 | 4.32 | 3.36 | 2.37 | 0.00 | 0.00 | 0.24 | 14.69 | 7.04 | 7.94 | 1 |
| NWA 6013 | I44 | Troilite | Sulfide | 32.91 | 0.00 | 5.23 | 0.00 | 6.07 | 22.62 | 0.00 | 0.00 | 0.22 | 19.66 | 13.29 | 1 |
| NWA 6013 | I46 | Labradorite | Feldspar | 55.76 | 1.04 | 1.11 | 7.64 | 11.90 | 0.00 | 3.26 | 0.00 | 0.00 | 0.64 | 18.65 | 1 |
| NWA 6013 | I47 | Pigeonite | Pyroxene | 57.44 | 0.00 | 8.46 | 0.30 | 13.78 | 0.00 | 0.64 | 0.00 | 0.14 | 3.93 | 15.31 | 1 |

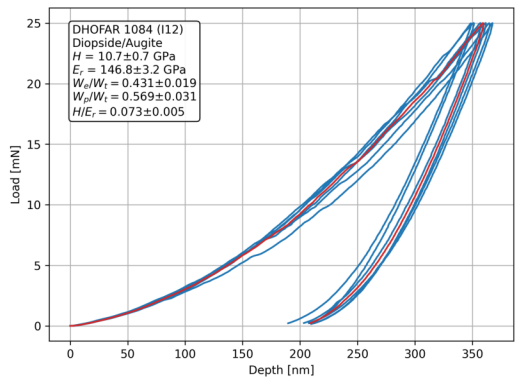
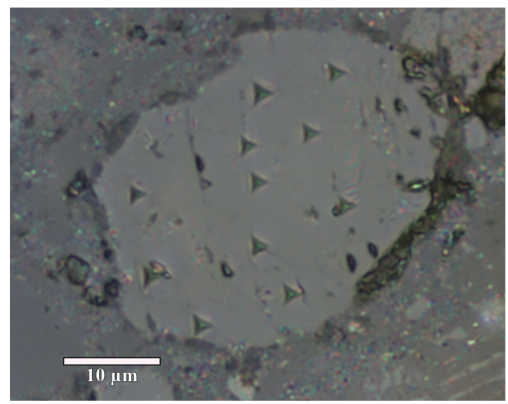
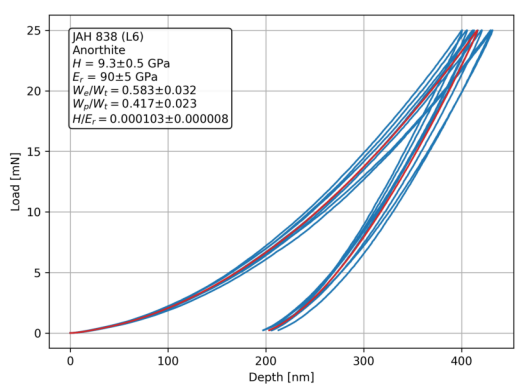
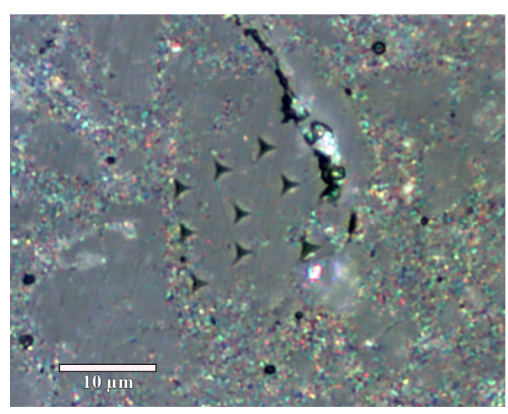
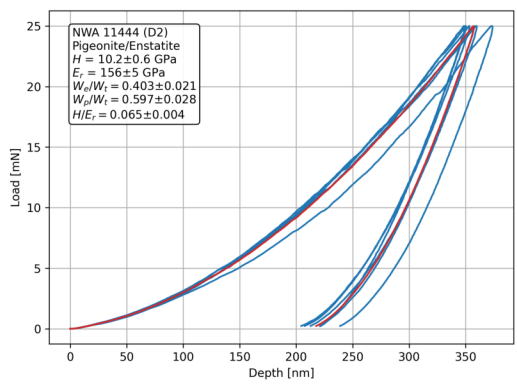
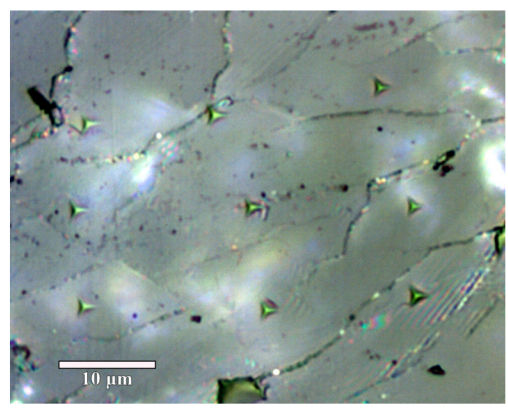
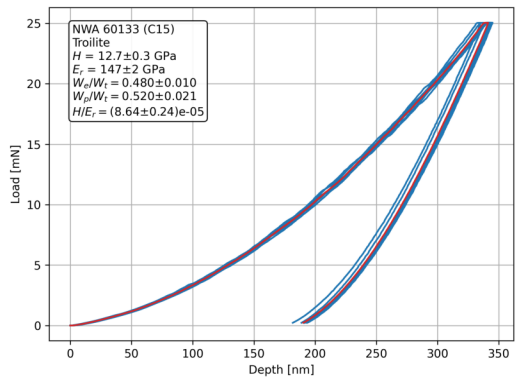
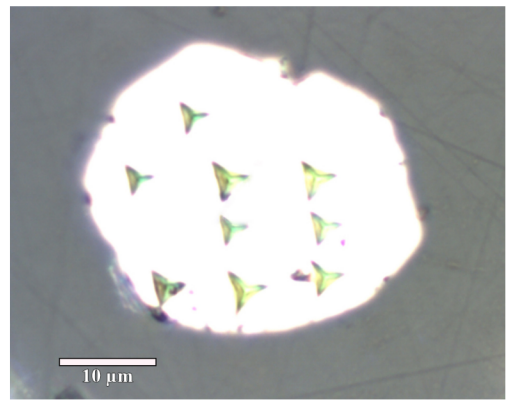

**Figure 6.** Representative nanoindentation curves and residual indentation imprints of different minerals in each meteorite. The median value is shown in red. Optical microscopic images of the indentation impressions are shown to the right of each curve. Indentations located on or near the edges of mineral phases have been systematically excluded.

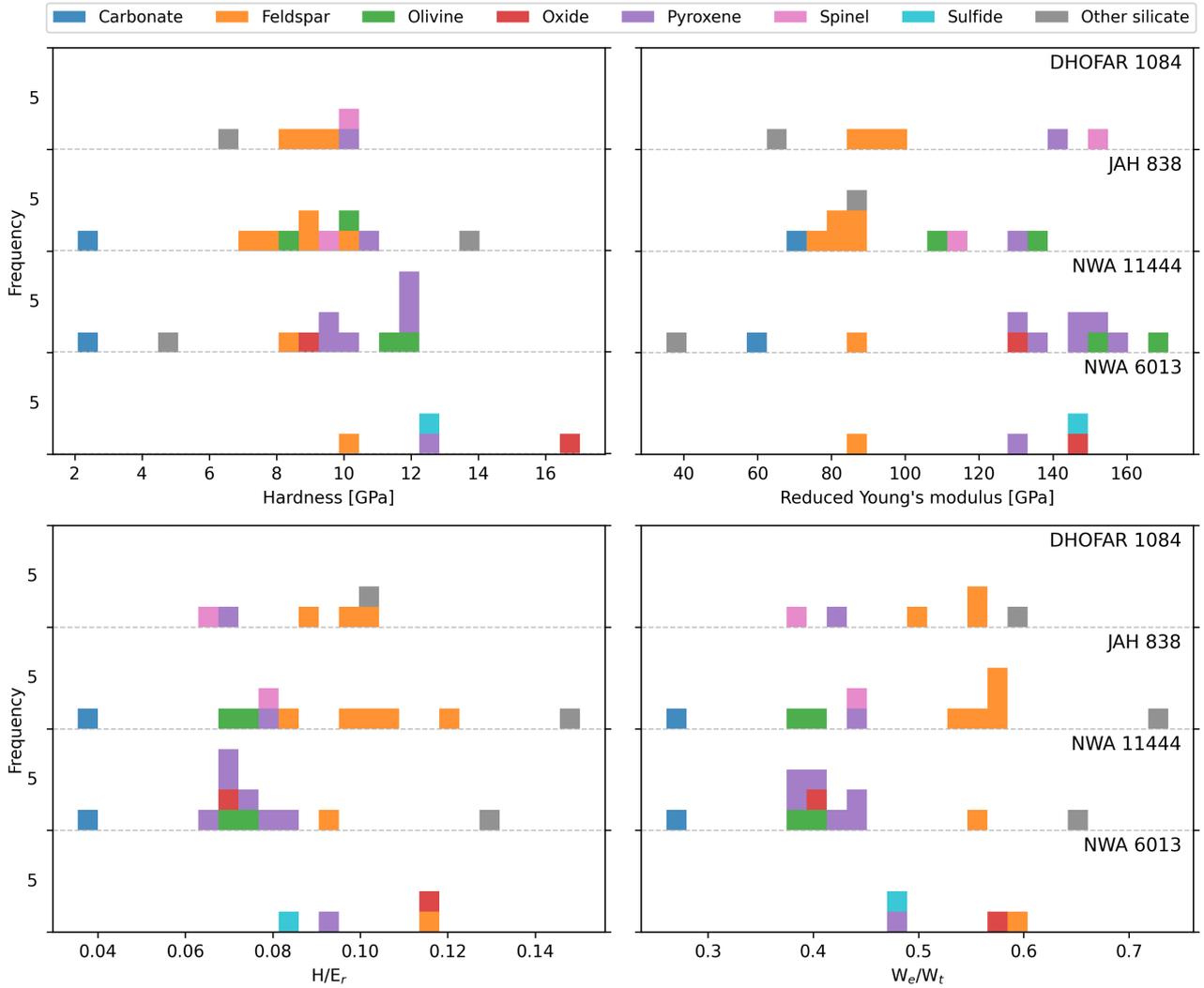

**Figure 7.** Histograms of hardness ($H$), reduced Young's modulus ($E_r$), $H/E_r$, and elastic recovery ($W_e/W_t$) for the meteorites examined in this study are presented in a single panel, with fixed frequency intervals of 5 to facilitate comparison.

**Figure 7** shows that lunar meteorites feature a similar distribution of hardness across all nanoindentation tests, featuring a predominant concentration around 9 GPa with a small secondary cluster at approximately 3 GPa. In contrast, the HED meteorite exhibits a distinct pattern: none of the tested minerals fall within this lower-hardness secondary group and contains the hardest mineral. Concerning the reduced Young's Modulus, both DHOFAR 1084 and NWA 11444 show comparable distributions, with the former encompassing the range's minimum and maximum values. In terms of the $H/E_r$ ratio and elastic recovery, JAH 838 and NWA 11444 demonstrate similarities, although the latter spans the minimum and maximum values within the dataset. For all examined properties except hardness, the HED meteorite, NWA 6013, displays intermediate values and a distribution that diverges modestly from those of the lunar meteorites, albeit the number of samples is smaller. It is pertinent to acknowledge that despite the presence of common mineralogical constituents in lunar meteorites, out of the four minerals identified within HED meteorites, three possess a unique mineralogy.

The upper panel of **Figure 8** portrays the average hardness of each mineral as a function of reduced Young's modulus, where each individual data point has been categorized by the respective meteorite and mineral group they belong to. Distinct mechanical properties emerge within each mineral group. Olivines, pyroxenes, feldspars, and spinels exhibit strikingly similar characteristics. Conversely, other silicate, and oxide minerals display significant variations in the mechanical properties. The carbonate group stands out as having the lowest hardness, while the oxides exhibit the highest hardness values.

The lower panel of **Figure 8** presents the average hardness of each mineral as a function of reduced Young's modulus. In this case, the data points have been categorized based on the identified mineral type. The three indented silicates (others) exhibit distinct mechanical characteristics. A notable correlation is evident in their atomic compositions: as the other element percentage increases, both hardness and Young's modulus show a corresponding increase.

Most of the identified minerals share similar compositions within a given group, with slight variations observed in certain elements. Based on the $H/E_r$ ratio, the mechanical properties within each mineral group generally exhibit consistency, except for the oxide and other silicate. However, there is observed variability in the mechanical properties of individual minerals. For instance, the other silicates display $E_r$ of 38, 66, and 90 GPa, respectively, while forsterites show values of 113, 139, and 174 GPa. Conversely, minerals such as anorthites and pigeonites present a more uniform values in their mechanical properties.

In indented regions where partial mixing with another mineral or a notable concentration of certain elements is detected, as seen in anorthite, diopside, and pigeonite, we observe no substantial variations in mechanical properties.

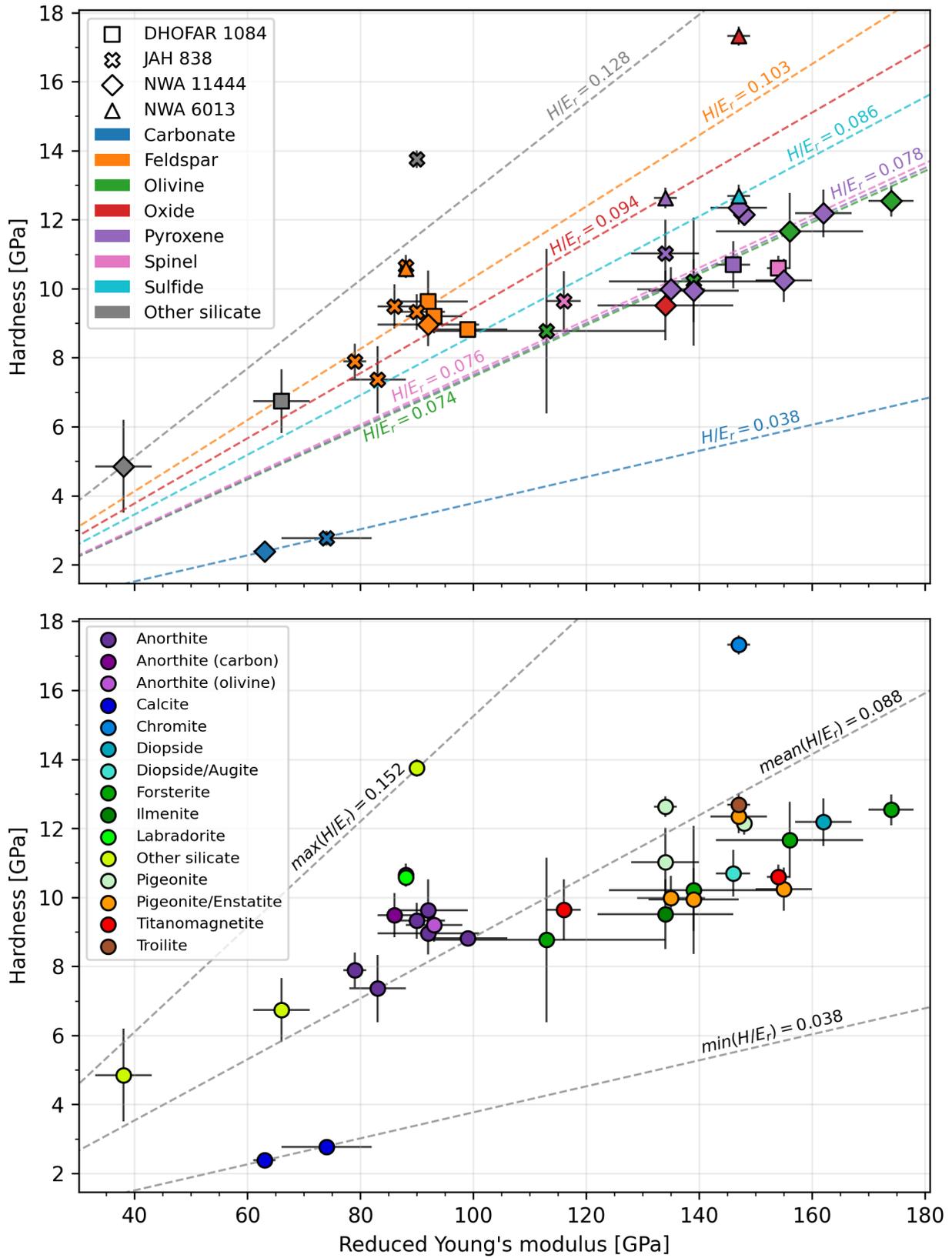

**Figure 8.** Mean hardness versus mean reduced Young's modulus for all the nanoindentations performed. The top panel shows meteorite and group classification with mean $H/E_r$ depicted for each group. The bottom panel shows mineral classification with global minimum, maximum, and mean. Error bars are depicted in black.

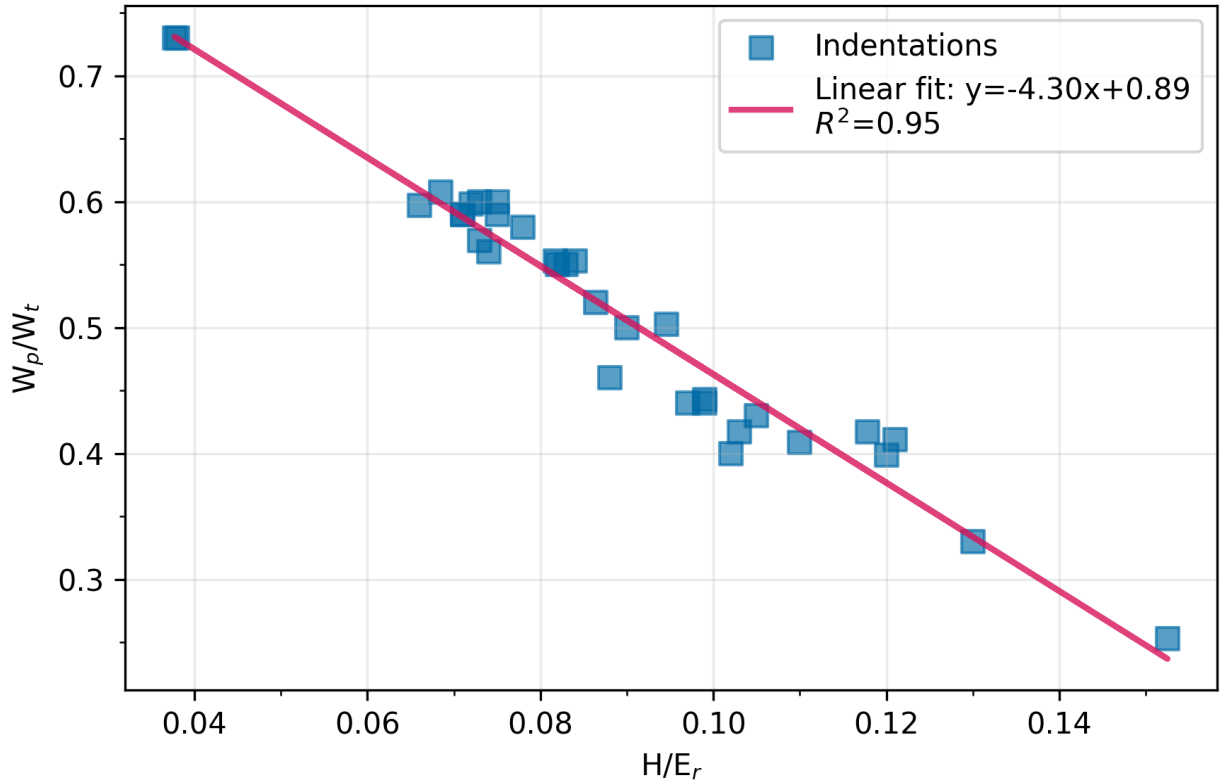

**Figure 9.** Scatter plot of $H/E_r$ ratio versus $W_p/W_t$ with square markers. The linear fit is depicted, along with its parameters and the coefficient of determination, highlighting the linear relationship between these properties.

**Figure 9** illustrated a distinct linear relationship between the $H/E_r$ and the normalized plastic energy ($W_p/W_t$), a trend that extends to the elastic component as well. This correlation is quantified through a linear regression analysis, the results of which are presented in the legend. The data points, represented by square markers, align closely with the fitted linear trend, signifying a strong empirical relationship. The coefficient of determination value ($R^2$=0.95) is indicative of the strength of this correlation.

## 4. DISCUSSION

The comprehensive examination of the mechanical properties of the lunar meteorites DHOFAR 1084, JAH 838, and NWA 11444 provide insights into various aspects of lunar science and ISRU. These meteorites from our celestial neighbor offer a unique window into the lunar landscape that human missions will encounter.

DHOFAR 1084 has been used to unravel the geologic history of the Moon (**Russell et al., 2004**). Chemical analyses of this meteorite have revealed a substantial abundance of aluminum and other refractory elements, strongly suggesting that it originates from the lunar crust. Furthermore, the presence of impact glass and brecciated fragments within DHOFAR 1084 indicates that it was likely formed through violent impact events—a common geological process that has left an indelible mark on the lunar surface over eons. The findings of this study underscore the significance of DHOFAR 1084 in shedding light on lunar geological processes.

JAH 838 has provided information on the chemical and isotopic composition of the Moon's interior. Researchers meticulously analyzed the isotopic ratios of oxygen, titanium, and other elements within this meteorite and discovered a striking resemblance to the samples brought back from the Apollo missions (**Korotev, 2017**). This congruence strongly confirms JAH 838's lunar origin. The chemical composition of JAH 838 also suggests its formation during the early stages of lunar history, a period marked by intense volcanic activity. JAH 838 is a fragmental breccia with mineral fragments of plagioclase, augite, pigeonite, poor-Ca pyroxene, and olivine set in a fine-grained matrix. Additionally, it contains lithic clasts of anorthosite, fragmental breccia, gabbro, norite, basalt, pyroxenite, and regolith (HASP spherule and chondritic lunar metal) embedded in a fine-grained dark gray matrix with metallic Ni-Fe, troilite, pyrrhotine, pentlandite, ilmenite, chromite, zirconolite, baddeleyite, and zircon (**Bouvier et al., 2017**).

NWA 11444 attests to the richness of lunar materials encapsulated within a fine-grain matrix (**Gattacceca et al., 2019**). The meteorite boasts an abundance of minerals and lithic clasts, featuring a diverse assortment of angular fragments, ranging from coarse-grained to aphanitic gabbros and basalts. Single-crystal types further diversify the composition of this meteorite, offering an idea of the lunar geological diversity. **Gattacceca et al. (2019)** delve into the remarkable attributes of NWA 11444, emphasizing its significance as a repository of lunar materials.

In turn, NWA 6013, the sole non-lunar meteorite under investigation in this study, is categorized as an olivine diogenite. These meteorites are believed to have originated at moderate depths within the mantle of the HED parent asteroid, which is thought to be Vesta. An interesting characteristic of NWA 6013 is the lack of clear boundaries between its subunits, imparting an overall metamorphic appearance to the rock. A noteworthy feature observed within NWA 6013 is the entrapment of olivine within pyroxene crystals, particularly in the pyroxene-rich subunits (**Ruzicka et al., 2015**). Additionally, the meteorite contains chromite grains, some of which attain sizes of several millimeters, further enriching its mineralogical diversity. Minor quantities of troilite and metallic iron are also present in the composition of this meteorite. In a broad sense, our present experiments indicate that NWA 6013 does not exhibit significantly distinct mechanical properties when compared to lunar meteorites.

Regarding the mean elemental composition measured in this study, each meteorite manifested a unique profile. DHOFAR 1084 was particularly rich in oxygen, aluminum, and titanium, whereas NWA 11444 excelled in magnesium and silicon concentrations. JAH 838 was noteworthy for its elevated calcium levels, which may have specialized chemical applications. NWA 6013, the sole HED meteorite analyzed, exhibited a compelling elemental composition, notably featuring the highest iron content among the samples. Moreover, it presented increased concentrations of chromium and sulfur. Overall, a notable variability was observed in both mechanical and elemental properties across all meteorites, suggesting material heterogeneity. Such variability highlights the imperative for additional investigations to elucidate the microscale spatial distribution of these attributes.

The calcite identified in JAH 838 and NWA 11444 lunar meteorites is attributed to terrestrial weathering (**Rubin, 1997**). This mineral is commonly observed as a fine-grained substance filling fractures linked with metal alteration products. It also appears as a surface coating on sections of meteorites that were initially buried in soil and in minor quantities occluding intergranular pores within weathered meteorites (**Al-Kathiri et al., 2005**). For instance, **Huidobro et al. (2021)** detected calcite in the lunar meteorite NWA 11273, which was discovered in the same region as NWA 11444.

The mineral chromite in the HED meteorite NWA 6013 demonstrates the highest mean hardness value at 17.3±0.3 GPa, indicating its superior resistance to deformation compared to the others. The particular case of olivine will be discussed in detail below. On the other end of the spectrum, the other mineral silicate in NWA 11444 has the lowest mean $E_r$ value at 38 GPa, implying more strain for a given constant stress. For comparison, the Chelyabinsk meteorite $E_r$ was determined to range from 69 to 78 GPa, as reported by **Moyano-Cambero et al., (2017)**. This contrasts with the values obtained for samples from Itokawa, which ranged from 82 to 111 GPa (**Tanbakouei et al., 2019**). The Young's modulus values for ordinary chondrites, as reported by **Yomogida & Matsui (1983)**, range from 10 to 140 GPa. In our study, the meteorite NWA 11444 exhibited the most variable $E_r$, with a minimum value of 38 GPa and a maximum of 174 GPa. The mean $E_r$, based on our indentation tests, was calculated to be 112±37 GPa.

The other mineral silicate (cell I30) in JAH 838 has the highest mean $H/E_r$ value of 0.1525, suggesting a relatively higher hardness compared to its elasticity. This might be indicative of a material with a higher wear resistance (**Pellicer et al., 2021**). Minerals from the oxide group, specifically chromite, shows varying mechanical properties, with chromite being the hardest, while other silicates have the highest $H/E_r$ ratio. The feldspar group, comprising anorthite and labradorite, generally shows moderate hardness and elasticity. While hardness and elasticity could be influenced by the mineralogical composition, they are not solely dictated by it.

While, in broad terms, minerals tend to exhibit analogous properties, the differences observed among them may arise from different factors, such as the presence of mixed second phases, phase insertions within the matrix, structural orientations, or porosity. Similarly, the varying degrees of shock evident in **Figures 1-4** may also suggest differences in the structural fragility of these phases. It is worth mentioning that the non-lunar meteorite examined (NWA 6013) for comparison in this study generally exhibits higher values of hardness, albeit with a smaller number of minerals tested.

A high $H/E_r$ ratio typically signifies a material that is relatively brittle and therefore easier to fracture, which is beneficial for mining operations. The meteorite JAH 838, rich in silicate minerals, exemplifies such a material, offering high hardness relative to its elasticity. For construction and manufacturing activities, materials with both high hardness and a high Young's Modulus are sought to ensure not only resistance to deformation but also good elastic recovery properties. The variability of the wear coefficient and its dependence on mechanical properties and abrasive geometry warrant further investigation for precise calibration in Moon ISRU and asteroid mining operations (**Pintaude, 2013**). Moreover, the material behavior during wear underscores the imperative for selecting mining tools composed of materials significantly harder than the target resources. Elastic effects also emerge as a critical factor in determining wear rates, suggesting that these considerations should be integrated into future predictive models to enhance the efficiency of resource extraction.

Mechanical properties like hardness and elasticity are often influenced by the material's mechanical history, including factors such as impacts, thermal history, and crystal structure. A meteorite subjected to high-impact events may have more refined crystal sizes due to shock-induced recrystallization, potentially enhancing its hardness and Young's modulus. Similarly, a meteorite that has undergone thermal annealing may display altered mechanical properties due to the depletion of moderately volatile phases (**Tartèse *et al.*, 2021**). Therefore, while the elemental composition provides some insights, the mechanical history of these meteorites likely plays a crucial role in defining their current mechanical properties. Further investigations, such as microstructural analyses or crystallographic studies, are essential to better understand these phenomena.

The linear correlation observed in our study between the $H/E_r$ ratio and normalized plastic energy resonates with the findings of **Varea *et al.* (2012)**. Their research on nanostructured Cu-rich CuNi electrodeposited films demonstrated a similar linear relationship in mechanical properties under nanoindentation. The agreement of the results not only validates our findings but also reinforces the reliability of the $H/E_r$ ratio as a robust parameter for predicting mechanical behavior in different material compositions and structures.

The investigation of the mechanical properties of minerals in lunar meteorites, conducted through indentation and other experimental methods, holds significant relevance for future ISRU. A primary objective of such processes is the utilization of extraterrestrial resources, including minerals, to fabricate structures and products capable of supporting human activities on the lunar surface. To achieve this objective, a comprehensive understanding of the mechanical behavior of rocks to be employed in manufacturing processes is crucial. Exploring the mechanical properties of lunar meteorites and their constituent minerals offers researchers valuable insights, enabling the extrapolation of these properties to the larger scale of lunar rocks, suitable for manufacturing. In this regard, **Tang et al. (2022)** emphasize the significant role of interphases between minerals and microcracks in determining rock mechanics. Such knowledge is fundamental for developing manufacturing processes and techniques adapted to lunar conditions and materials.

Insights garnered from studying the mechanical properties of minerals like olivine in lunar meteorites can inform the creation of manufacturing processes for producing robust structural materials and components, characterized by strength, durability, and resistance to wear. Additionally, the understanding of lunar materials can drive the development of novel alloys and composites optimized to thrive in the lunar environment and cater to the specific demands of lunar manufacturing. Enhanced comprehension of the mechanical properties of lunar materials facilitates multiple facets of future colonization activities, encompassing material characterization for quality control, coating design, and optimization, wear and fatigue assessment, and the development of new materials, among others.

Porosity, or the presence of voids within a material, plays a pivotal role in shaping its mechanical properties. Porosity can influence strength, stiffness, fracture toughness, and resistance to fatigue and wear. For instance, experimental measurements have demonstrated that an increase in porosity of approximately 20% results in a reduction of silicon's Young's modulus by nearly 50% (**Magoariec & Danescu, 2009**). Lunar regolith, the surface layer of loose material on the Moon, exhibits notable porosity, ranging from approximately 4% to 21%, with an average of around 12% (**Wieczorek et al., 2013**). This substantial porosity poses challenges when considering the direct use of lunar regolith as a construction material, as it may lack the necessary strength and stability. In **Figure 8**, varying mechanical properties across different phases are observed, yet align along a constant $H/E_r$ ratio. This alignment could be associated with changes in local porosity or density, indicating that as the porosity of a mineral increases, its properties diminish along the axis of a constant $H/E_r$ ratio (**Luo & Stevens, 1999**).

**Kumamoto et al. (2017)** reported that Young's modulus for terrestrial olivines ranges from 160 to 290 GPa. In contrast, our measurements for lunar olivines exhibit a range from 113 to 174 GPa. Correspondingly, hardness values for lunar olivines are lower, between 8.8 and 12.5 GPa, as opposed to terrestrial counterparts which range from 14 to 18 GPa. This disparity in mechanical properties compared to terrestrial olivines—irrespective of whether they are single crystals or polycrystalline aggregates—underscores the necessity for comprehensive characterization of lunar materials for ISRU, given their distinct behavior. The surface materials of the Moon have been subjected to numerous impacts, necessitating substantially more forceful impacts to eject these lunar rocks into orbit. This could potentially result in lunar meteorites that are structurally weakened, exhibiting increased porosity or microfracturing. While porosity in olivine may predominantly arise from impact-induced microfracturing, we cannot discount the possibility of vaporization-induced voids in the crystalline structure during the shock wave transit.

To address this challenge, one approach involves the development of techniques for compacting and sintering lunar regolith to produce denser, stronger construction materials. Nanoindentation testing emerges as a valuable tool within this process, offering a non-destructive means of assessing the mechanical properties of compacted regolith samples, including hardness, elasticity, and fracture toughness.

In addition, the utilization of micromechanical properties derived from indentation testing holds immense potential for advancing our understanding of material behavior and enhancing the accuracy of impact models. By considering the complex responses of materials at the microscale, we can refine simulations, optimize designs, and better predict how materials perform under high-velocity impacts.

## 5. CONCLUSIONS

Lunar meteorites, originating from our celestial neighbor, the Moon, have captivated the scientific community for their intrinsic value in unraveling the mysteries of lunar geology and history. Ejected from the lunar surface through the tumultuous forces of high-velocity impacts, these enigmatic rocks make their journey to Earth, offering an unparalleled window into the composition, properties, and shock experiences of lunar materials. In contrast to the Apollo missions, where samples were meticulously collected from specific lunar locales, lunar meteorites represent fragments of the Moon with origins veiled in uncertainty and, in principle, randomness.

This study has provided a comprehensive analysis of the mechanical and elemental properties of lunar meteorites DHOFAR 1084, JAH 838, NWA 11444, and HED meteorite NWA 6013. The findings reveal significant heterogeneity in both mechanical and elemental attributes across the samples, underscoring the need for further investigation into the microscale spatial distribution of these properties.

The majority of the minerals identified in this study, particularly olivines, feldspars, pyroxenes, and spinels, demonstrate closely aligned compositional and mechanical characteristics. On the other hand, other silicate and oxide minerals exhibit important variations in their mechanical properties. The carbonate group distinguishes itself by manifesting the lowest hardness values, whereas the oxides are characterized by the highest hardness.

Forsterite in NWA 11444 has the highest reduced Young's Modulus value. In contrast, other silicate in the same meteorite shows the lowest mean Reduced Young's Modulus. The three indented silicates (others) exhibit distinct mechanical characteristics, which not only have implications for their potential applications but also raise questions about the underlying factors influencing these properties.

Olivines from terrestrial origins, when analyzed using nanoindentation techniques, demonstrate greater hardness and a superior Young's modulus relative to olivines of lunar origin. A linear correlation is observed between the $H/E_r$ ratio and normalized plastic and elastic energies. Furthermore, various mineral phases display a constant $H/E_r$ ratio axis, which may indicate variations in local porosity. The need for further studies is evident, particularly focusing on aspects such as porosity, phase insertions within the matrix, and structural orientations, to provide a more comprehensive understanding of these mechanical characteristics.

By comprehending the mechanical characteristics of these compacted samples, researchers can gain valuable insights into optimizing compaction and sintering processes to yield more robust and durable building materials for lunar applications and asteroid mining. This advancement lays the groundwork for the creation of sustainable infrastructure, encompassing habitats, roadways, and other vital structures essential for long-term human presence on the Moon.

The use of microscale rock mechanics experiments to characterize the mechanical properties of minerals in meteorites provides valuable insight into the composition and behavior of materials on the lunar surface. The data obtained can be used to guide the development of extraterrestrial manufacturing techniques and space exploration technology, as well as contribute to next-generation impact modeling studies. The combination of mineral identification and nanoindentation testing provided a comprehensive characterization of the mechanical properties of the meteorites studied, which can inform forthcoming research efforts in space science and ISRU.

For future research, we aim to conduct a comparative analysis of the mechanical properties of lunar rocks, incorporating both Apollo mission return samples and analogous terrestrial minerals.

*Acknowledgments*- JMT-R and EP-A acknowledge financial support from the project PID2021-128062NB-I00 funded by MCIN/AEI/10.13039/501100011033. The lunar samples studied here were acquired in the framework of grant PGC2018-097374-B-I00 (P.I. JMT-R). This project has received funding from the European Research Council (ERC) under the European Union's Horizon 2020 research and innovation programme (grant agreement No. 865657) for the project "Quantum Chemistry on Interstellar Grains" (QUANTUMGRAIN). AR acknowledges financial support from the FEDER/Ministerio de Ciencia e Innovación – Agencia Estatal de Investigación (PID2021-126427NB-I00, PI: AR). Partial financial support from the Spanish Government (PID2020-116844RB-C21) and the Generalitat de Catalunya (2021-SGR-00651) is acknowledged. The authors extend sincere gratitude to the anonymous referee and to Jeffrey Wheeler for their valuable review, which have significantly enhanced the quality of this work.